\documentclass[conference]{IEEEtran}
\usepackage[cmex10]{amsmath}
\usepackage{amssymb}
\usepackage{graphicx}
\usepackage{subcaption}
\usepackage{fancyhdr}
\usepackage{bbm}
\usepackage{physics}
\usepackage[style=ieee]{biblatex}
\usepackage{amsthm,amsfonts,paralist,mathtools,physics}
\usepackage[hidelinks]{hyperref}
\usepackage{color,dsfont} 
\makeatletter
\renewcommand*\env@matrix[1][*\c@MaxMatrixCols c]{%
  \hskip -\arraycolsep
  \let\@ifnextchar\new@ifnextchar
  \array{#1}}
\makeatother

\usepackage[normalem]{ulem}
\usepackage{balance}

\usepackage{listings}
\renewcommand\refname{}

\newcommand{\floor}[1]{\lfloor #1 \rfloor}

\newcommand{\tensor}{\otimes}

\newcommand{\GF}{\text{GF}}
\newcommand{\qcode}[1]{\left[\!\left[ #1 \right]\!\right]}

\newcommand{\F}{\mathbb{F}}
\newcommand{\N}{\mathbb{N}}

\theoremstyle{plain}
\newtheorem{remark}{Remark}
\newtheorem{example}{Example}

\allowdisplaybreaks[4]

\title{Fault-Tolerant QLDPC Syndrome Measurement via LDGM Encoding}
\author{
	\IEEEauthorblockN{Eren Guttentag\IEEEauthorrefmark{1}\IEEEauthorrefmark{2} and  Anthony G\'omez-Fonseca\IEEEauthorrefmark{3}}
	\IEEEauthorblockA{\IEEEauthorrefmark{1}Duke Quantum Center, Duke University, Durham, NC 27701, USA \\ \IEEEauthorrefmark{2}Department of Electrical and Computer Engineering, Duke University, Durham, NC 27708, USA \\ \IEEEauthorrefmark{3}Department of Mathematics and Statistics, University of South Florida, Tampa, FL 33620, USA}
}
\begin{document}
	\maketitle


\begin{abstract}

Quantum data-syndrome (QDS) codes are a class of quantum error-correcting (QEC) codes that allow for protection against errors both on the data qubits and on the syndrome itself via redundant measurement of stabilizer group elements. One way to define a QDS code is to choose a syndrome measurement (SM) code, a classical block code that encodes the syndrome of the underlying quantum code by defining additional stabilizer measurements to be performed. 
The properties of a chosen SM code affect the properties of these additional stabilizer measurements. Notably, the weight of the measured stabilizers depends on the weight of the columns of the generator matrix of the SM code. If the underlying quantum code is a quantum low-density parity-check (QLDPC) code, then the $O(1)$ weight of the stabilizer generators is essential for the advantageous properties of a QLDPC code, but a weight-$w$ column of a chosen SM code will result in stabilizer measurements on $O(w)$ qubits. For a SM code with unbounded column weight, this results in unbounded stabilizer weight, losing the QLDPC property.

To prevent this, we propose the use of certain low-density generator-matrix (LDGM) codes as SM codes. We use an efficient progressive-edge-growth-like algorithm to create LDGM SM codes with column and row weights that result in measured stabilizers that have constant weight, thus preserving the desirable properties of the underlying QLDPC code. This process allows for control over stabilizer weights and SM code distance.
This control results in significantly better performance than repeated syndrome extraction, allowing for both higher distances and fewer syndrome measurements. As syndrome measurements performed are a reasonable metric for the time a circuit takes to implement, we conclude that these LDGM codes allow for improved implementation of QLDPC codes without sacrificing the low weights of the syndrome measurements performed.

\end{abstract}

\section{Introduction}
In order to build a functional and scalable quantum computer, it is necessary to be able to accurately detect, identify, and correct errors on the constituent qubits \cite{Nielsen2010,Preskill2018}. This is typically accomplished by the use of a quantum stabilizer code, which is defined by the generators of an abelian group of Pauli operators \cite{gottesman1997}. The measurement of the group generators produces a binary syndrome that indicates the locations of errors. These measurements involve several multi-qubit gates, which can introduce errors on the qubits involved and corrupt the measurement outcome. The propagation of errors can be curtailed by minimizing the number of quantum gates being performed, while errors on the measurements are corrected by acquiring redundant syndrome information. One way that quantum gates are minimized is by the use of stabilizer codes whose generators involve a small ($O(1)$) number of qubits; such codes are known as quantum low-density parity-check (QLDPC) codes \cite{Breuckmann2021}.

A competing constraint on fault-tolerant and scalable quantum computing is the fact that many physical qubits are unstable and inevitably decohere due to interaction with the environment \cite{Nielsen2010}. Therefore, it is also desirable to perform all operations on the qubits as quickly as possible in order to avoid logical errors caused by decoherence. The strategy discussed in this paper is an attempt to balance these factors by obtaining syndrome information as efficiently as possible while constraining the number of quantum gates involved. 

To perform fault-tolerant syndrome extraction we use codes known as quantum data-syndrome (QDS) codes; such codes were introduced by Fujiwara \cite{Fujiwara2014,Fujiwara2015} and Ashikhmin et. al \cite{Ashikhmin2020,Ashikhmin2014,Ashikhmin2016}. The particular QDS code construction we discuss uses classical error correction codes as syndrome measurement (SM) codes \cite{Guttentag23} in order to define an overdetermined set of stabilizers to measure. In particular, we will construct codes that have low-weight columns. 

In order to show the value of such codes, we consider a phenomenological error model in which qubit errors manifest before the application of syndrome measurements, and measurement errors occur to flip the result of measured syndrome bits. We use a technique known as importance sampling \cite{Gutierrez2019,Guttentag23} to perform these simulations efficiently.

 \section{Background}
    \subsection{Stabilizer codes}
    
	Consider the Pauli matrices $P_0=I=\smqty[1&0\\0&1]$, $P_1=X=\smqty[0&1\\1&0]$, $P_2=Y=\smqty[0&-i\\i&0]$, and $P_3=Z=\smqty[1&0\\0&-1]$. The Pauli operators on $n$ qubits are $n$-fold tensor products of Pauli matrices of the form \begin{equation*}
    i^{c}\cdot P_{a_0}\tensor P_{a_1}\tensor\cdots\tensor P_{a_{n-1}},
    \end{equation*}
    with $a_i,c\in\{0,1,2,3\}$. For the sake of simplicity, we will omit tensor products from our notation; so the three-qubit operator $X\tensor Y\tensor I$ becomes $XYI$, for example. These operators form the group $\mathcal{P}^n$ of $n$-qubit Pauli operators.
    
    An $\qcode{n,k,d}$ stabilizer code encoding $k$ logical qubits into $n$ physical qubits is defined by its \textit{stabilizer generators}, a set of $\ell:=n-k$ independent and commuting Pauli operators which we denote $\{g_1,g_2,\dots,g_\ell\}\subset\mathcal{P}^n$. These operators generate the stabilizer group $\mathcal{S}\subset\mathcal{P}^n$ of the code, which is commutative, does not contain $-I^{\tensor n}$, and has order $2^{\ell}$. These are called stabilizers because the valid codewords of the quantum code are the joint $+1$ eigenstates of the stabilizer group, and are thus stabilized by it: for any stabilizer $g_i\in\mathcal{S}$ and any codeword $\ket{\psi}$ of the associated quantum code, $g_i\ket\psi=\ket\psi$.
    
    A Pauli operator on $n$ qubits can be seen as a length-$n$ vector with entries in $\GF(4)$, under the homomorphism $\tau:\mathcal{P}\to\GF(4)$ that maps $I\to0$, $X\to1$, $Y\to\bar\omega=1+\omega$, and $Z\to\omega$, and ignores any global phase of $\pm1,\pm i$. This can be naturally extended to send operators on $n$ qubits from $\mathcal{P}^n$ to $\GF(4)^n$ \cite{Ashikhmin2020}. Under this homomorphism, multiplication of stabilizers corresponds to bitwise addition in $\GF(4)^n$. We will use $g\in\mathcal{P}^n$ to refer to a stabilizer and $\mathbf{g}\in\GF(4)^n$ to refer to the corresponding length-$n$ $\GF(4)$-vector $\tau(g)$. 
    
    In general, an error $E$ on an $\qcode{n,k,d}$ quantum code can be represented as an element of $\mathcal{P}^n$. To determine what error is occurring, we traditionally measure the stabilizer generators \cite{Nielsen2010,gottesman1997}. The measurement of a code's $\ell$ stabilizer generators gives as an output a length-$\ell$ binary vector called the syndrome $\mathbf{s}=(s_1,s_2,\dots,s_\ell)$. The $i$-th syndrome bit $s_i$, corresponding to the stabilizer operator $g_i$, is 0 if $E$ and $g_i$ commute, and 1 if they anticommute. For vectors $\mathbf{x},\mathbf{y}\in\GF(4)^n$ with elements $x_i,y_i\in\GF(4)$, the analogous function is the \textit{trace inner product} $\mathbf{x}\star\mathbf{y}:\GF(4)^n\cross\GF(4)^n\to \GF(2)$:
    \begin{equation*}
\mathbf{x}\star\mathbf{y}=\sum_{i=1}^n(x_i\bar{y_i}+\bar{x_i}y_i),
    \end{equation*}
    where $\bar{0}=0$, $\bar{1}=1$, $\bar{\omega}=1+\omega$, and $\overline{1+\omega}=\omega$, and multiplication in GF(4) is the usual  \cite{Calderbank1997,Guttentag23}. Essentially, the $i$-th element of the sum is 0 if $x_i=0,\,y_i=0$, or $x_i=y_i$, and 1 otherwise. The trace inner product is 0 when the Pauli operators represented by $\mathbf{x}$ and $\mathbf{y}$ commute, and is 1 when they anticommute, serving as a quantum generalization of a classical parity check.

	\subsection{Effective Distance}
    A characteristic problem in quantum error correction (QEC) is the fact that errors that occur on ancilla qubits that propagate to the data qubits through the use of two-qubit gates (often CNOTs) \cite{Dennis_2002}. With a single ancilla qubit per stabilizer, it is possible for an error to occur on that ancilla qubit and then propagate through every gate onto data qubits. For a stabilizer with weight $w$, this can result in $w$ errors introduced to the data qubits. In the worst case, this results in the effective distance of the code being lowered to $\tilde{d}=\floor{\frac{d}{w}}$, as $\floor{\frac{t+1}{w}}$ errors during the extraction process can result in $t+1$ errors being introduced on the code qubits (which in the right order can cause a logical error) \cite{Dennis_2002,golowich2023}.

    Such errors are therefore propagated depending on the weight of the stabilizer being measured. While individual hook errors are unlikely, it is still best to protect against the possibility of these errors by minimizing the weight of the stabilizer group elements being measured \cite{Bombin2015}. This paper eventually aims to show that it is possible to perform fault-tolerant syndrome extraction without asymptotically lowering the effective distance of the code.

   \subsection{LDPC and QLDPC codes} 

    A class of binary error correction codes called \textit{low-density parity-check} (LDPC) codes were proposed by Gallager \cite{Gallager62,Gallager63}. A binary code is LDPC if the number of nonzero entries (or Hamming weight) in each row and column of the parity-check matrix are both bounded above by a relatively small constant. As a consequence, LDPC codes have sparse parity-check matrices and sparse Tanner graphs.

    As with many classical error correction codes, the theory of LDPC codes has been extended into quantum error correction codes. Specifically, quantum LDPC (QLDPC) codes are families of quantum stabilizer codes whose stabilizer generators satisfy the LDPC property. This means that each stabilizer has a constant number of non-identity Pauli operators and each qubit is in the support of a constant number of stabilizers \cite{Breuckmann2021}. Families of these QLDPC codes are those where this constant bound does not change when the distance of the code is increased; one example of this is the family of Kitaev surface codes \cite{freedman1998}.

    When implementing quantum codes in practice, this QLDPC property reduces the weight of errors that propagate due to hook errors \cite{Breuckmann2021,camara2005,golowich2023}. In fact, under the definition of effective distance given in Section II.B, this allows for the effective distance $\overline{d}$ to continue to scale with the distance of the quantum code as the size of the code increases: $\overline{d}\in O(d)$. These QLDPC codes allow for stabilizer measurements to be performed with circuits of constant depth \cite{Tremblay2022}. 

\section{Data-Syndrome Codes}

    \subsection{Protecting syndrome information}
    \label{subsec:protecting_syndrome_information}
    We assume here that syndrome information is obtained by the use of \textit{Shor-style} syndrome extraction, in which a weight-$w$ stabilizer can be measured fault-tolerantly using $w$ gates \cite{Shor1996,DiVincenzo1996}. It is possible to do this using $w$ ancilla qubits, or a single ancilla qubit, depending on the error model of the system being worked with.
    
    Because syndrome information is obtained by using ancilla qubits, it is possible for Pauli errors to occur on those qubits just as they can on data qubits. We are measuring a classical bit for each stabilizer element, so we only consider these errors that manifest as a bit flip on these classical bits. These are called \textit{measurement errors} and can be considered to be classical errors on the classical syndrome. For a more in-depth discussion of these errors and their consequences, see \cite{Ashikhmin2020,Fujiwara2014}.

   When measurement errors can occur, it is important to distinguish between the \textit{correct} syndrome $\mathbf{s}$ and the \textit{measured} syndrome $\hat{\mathbf{s}}$. They are related by
   \begin{equation*}
   \hat{\mathbf{s}}=\mathbf{s}+\hat{e},
   \end{equation*}
   where $\hat{e}\in \F_2^n$ is a binary \textit{error vector} such that the $i$-th bit of $\hat{e}$ is 0 if there is no measurement error on the syndrome bit $s_i$, and 1 if a measurement error occurs. Note that the sum is bitwise in $\F_2^n$.
    
    We assume that each syndrome bit has equal probability $p_m$ of experiencing a measurement error. In practice, it is often the case that the probability of measurement error on a particular syndrome bit is proportional to the weight of the stabilizer being measured. In this paper, we will consider stabilizers of constant weight and therefore our assumption is sound.

    \subsection{Syndrome measurement codes}

    One way to define a suitable set of redundant stabilizer measurements to perform is by the use of a \textit{syndrome measurement} (SM) code. An SM code is a classical error-correcting code that is used to encode the syndrome information. For an $\qcode{n,k,d}$ quantum stabilizer code with $\ell=n-k$ stabilizers, an $[n_{SM},\ell,d_{SM}]$ classical code can be used to protect the stabilizer information against bit-flip measurement errors. Such a classical code has a binary generator matrix $G\in \F^{n_{SM}\times\ell}_2$. We use this generator matrix to define a set of stabilizer group elements to measure. Consider a quantum code with stabilizer generators $S_i$, $1\leq i\leq \ell$, and a classical code with an $n_{SM}\times \ell$ generator matrix $G$. Let the $i$-th row of $G$ be given as $\mathbf{r}^{(i)}\in\F_2^{n_{SM}}$ and denote the $j$-th element of that row be $r_j^{(i)}$. Note that $r_j^{(i)}\in\{0,1\}$.

    Then this code defines $n_{SM}$ syndrome measurements to perform, where the $k$-th syndrome group element to measure, $\hat{S}_k$, is defined as
\begin{equation}\label{eq:1}\hat{S}_k=\prod_{i=1}^\ell {r}^{(i)}_k S_i.\end{equation}

    An example will make this clearer. Consider a quantum code defined by three stabilizer generators, which we call $S_1$, $S_2$, and $S_3$, and let $s_i$ be the corresponding syndrome bit of $S_i$. We can choose a classical error-correcting code with three logical bits in order to encode this syndrome information. Let us choose the $[7,3,4]$ Hamming code with generator matrix $G$ given by
    \begin{equation*}
    G=\mqty[1&0&0&1&0&1&1\\0&1&0&1&1&0&1\\0&0&1&0&1&1&1].
    \end{equation*}
    Then this defines seven stabilizer group elements $\left\{\hat{S}\right\}$ to be measured. Equation \eqref{eq:1} tells us that the $i$-th stabilizer group element to be measured is the product of the stabilizers corresponding to nonzero entries in the $i$-th column of $G$, as shown in Table \ref{tab:enc}.
    
    \begin{table}[!h!]
        \centering
        \begin{tabular}{c|c|c}
            $i$& Corresponding column of $G$ & $\hat{S}_i$\\\hline
             1&  $\mqty[1&0&0]^T$& $S_1$\\
             2&  $\mqty[0&1&0]^T$& $S_2$\\
             3&  $\mqty[0&0&1]^T$& $S_3$\\
             4&  $\mqty[1&1&0]^T$& $S_1S_2$\\
             5&  $\mqty[0&1&1]^T$& $S_2S_3$\\
             6&  $\mqty[1&0&1]^T$& $S_1S_3$\\
             7&  $\mqty[1&1&1]^T$& $S_1S_2S_3$\\ 
        \end{tabular}
        
        \caption{When using the [7,3,4] Hamming code as an SM code on three stabilizers, seven stabilizer group elements are defined and measured. Each stabilizer measured corresponds to a column of the encoding matrix.}
        \label{tab:enc}
    \end{table}
    
    A key observation is that the syndrome bit $\hat{s}_i$ of a stabilizer group element $\hat{S}_i$ is the parity of the syndrome bits corresponding to the generators that define it. For example, in Table \ref{tab:enc}, the stabilizer group element $\hat{S}_7$ is given by the product $S_1S_2S_3$, and the syndrome bit $\hat{s}_7$ produced upon measuring this stabilizer is given by $\hat{s}_7=s_1\oplus s_2\oplus s_3$, where $\oplus$ indicates addition modulo 2. This tells us then that the length-$n_{SM}$ syndrome obtained by measuring the $n_{SM}$ stabilizer group elements without any measurement errors should be exactly the same as the encoding of the length-$\ell$ syndrome in the generator matrix $G$, which should be readily apparent. Therefore, after performing measurements to obtain a noisy length-$n_{SM}$ classical syndrome, a decoder of the classical SM code can be used to determine the length-$\ell$ encoded information, which will correspond to a valid syndrome of the underlying quantum code.

    \subsection{Stabilizer weights}

    Note that, as shown in Table \ref{tab:enc}, each column of the encoding matrix $G$ of the SM code defines a stabilizer group element to be measured. Moreover, the Hamming weight \cite{Nielsen2010} of the corresponding column of $G$ is equal to the number of stabilizer generators that are multiplied together to obtain the measured stabilizer. For a class of SM codes with unbounded column weights, this means that we will have to measure the product of arbitrarily many stabilizer generators. To see why this is disadvantageous, consider an arbitrary QLDPC code $\mathcal{C}$. Its stabilizer generators have a constant weight $O(1)$; we can call this weight $w_\mathcal{C}$. Then any product of $m\in\N$ stabilizers has a maximum weight of $m w_{\mathcal{C}}$\footnote{If the stabilizers involved in one such product overlap in their support, the weight can be less than this upper bound. However, as the distance of a QLDPC code increases, overlaps become increasingly unlikely. Therefore, we consider it reasonable to assume that most stabilizers measured will have weight close to this upper bound based on the corresponding column weight.}. 

    Therefore, the maximum column weight of the generator matrix has a linear relationship with the maximum weight of the stabilizers being measured. The effect of this is that a distinct advantage of the use of QLDPC codes---the low-weight stabilizers which allow for confinement of errors \cite{Bombin2015}---is lost when using syndrome measurement codes which have unbounded column weights. A high-weight column corresponds to the product of a large number of stabilizer operators, which will be a high-weight element of the stabilizer group, requiring a very large number of error-prone gates to implement.
    
    The current best-known construction for SM codes uses BCH codes to encode $\ell$ syndrome measurements of a code with distance $2t+1$ in $\ell+O(t\log(\ell))$ total syndrome measurements \cite{Guttentag23}. BCH codes can exhibit exactly this undesirable behavior: their column weights are unbounded. This can be seen from the fact that a $[2^r-1,2^r-r-1,3]$ binary Hamming code is a special case of a BCH code, and necessarily contains a column whose entries are all $1$, and therefore a column with maximal weight \cite{Hamming50,Bose1960a,Bose1960b,Hocquenghem1959}.
    
    Therefore, if it is desirable to preserve the QLDPC property of a code and still use a syndrome measurement code to define stabilizers to be measured, we want to be able to guarantee that the generator matrix of our SM code will have low-weight columns. We can achieve this by choosing \textit{low-density generator-matrix} (LDGM) codes. The process of choosing these codes and applying them as syndrome measurement codes is the focus of this paper.

    \section{The $\qcode{25,1,5}$ Rotated Surface Code}

\begin{figure}
        \centering
        \includegraphics[width=0.8\linewidth]{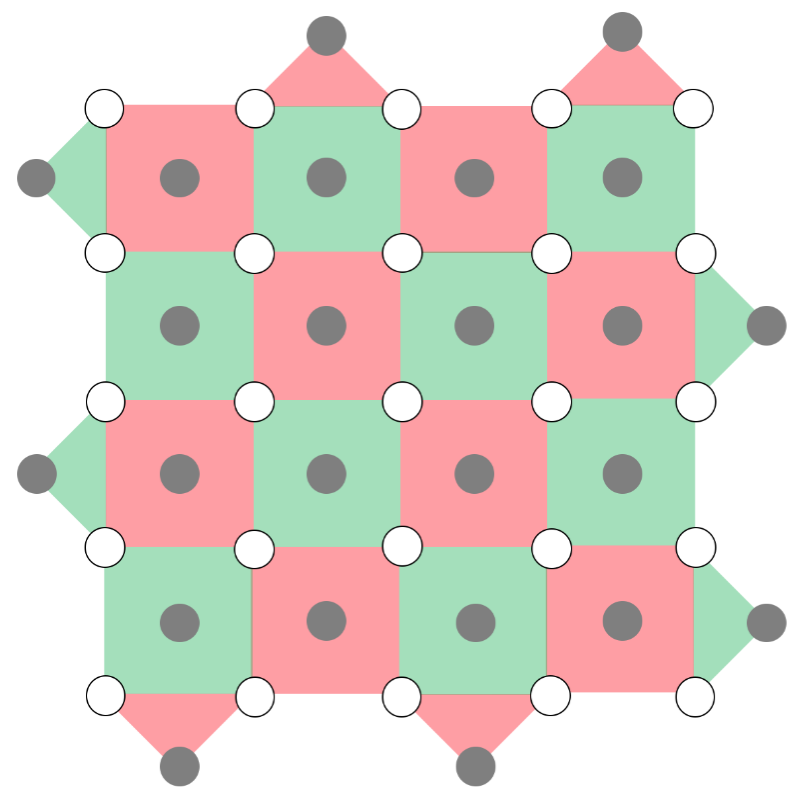}
        \caption{The distance-5 rotated surface code, illustration from \cite{Chamberland2018}. White dots correspond to the 25 data qubits, which we number sequentially as 1 to 25 in lexicographic order, while gray dots correspond to the ancilla qubits which measure their corresponding stabilizers. Green squares correspond to $Z$-stabilizers while red squares correspond to $X$-stabilizers.}
        \label{fig:rsc}
    \end{figure}

    To construct a suitable SM code, we first choose the underlying quantum code whose syndrome information we are attempting to extract. In this paper, as an illustrative example, we choose the $\qcode{25,1,5}$ 25-qubit \textit{rotated surface code}, illustrated in Fig. \ref{fig:rsc}. This code is a QLDPC code that encodes 1 logical qubit in 25 physical qubits, and has 24 ancilla qubits. Logical operators of this code are $\bar{Z} = Z_1Z_2Z_3Z_4Z_5$, a horizontal line, and $\bar{X} = X_1X_6X_{11}X_{16}X_{21}$, a vertical line, as well as any product of stabilizers with these logical operators. For a fuller description of these logical operators, see \cite{Chamberland2018,Tomita2014}. Note that all stabilizers are supported on at most 4 data qubits, and similarly that all data qubits are in the support of at most 4 stabilizers. This holds true for any size of rotated surface code, and is what allows these codes to be QLDPC.

    Because this code has distance 5, it can correct any error on up to 2 data qubits. When using this code, we would like it to be \textit{globally} 2-error correcting, i.e. able to correct any error on up to 2 data or ancilla qubits. Traditionally, this involves performing 5 rounds of syndrome extraction; with 24 ancilla qubits, this comes out to a total of 120 total stabilizer measurements. To show that a more concise system of syndrome extraction can be done, we decide that we want our syndrome extraction to be done in only 60 total stabilizer measurements, cutting the number of measurements in half\footnote{This decision is somewhat arbitrary; an individual choosing to adapt this syndrome measurement technique may choose a larger or smaller number of desired measurements to perform, depending on the particular errors they expect to see.}. Another reason we choose to perform 60 total measurements is that it has many shared factors with 24, the number of stabilizer generators of our code. The advantage of this is that it allows us to construct many different protographs, as will be explored in Section \ref{sec:ldgm_sm_codes_construction}.

    Additionally, before constructing a suitable LDGM code for our purposes, it is necessary to determine a maximum stabilizer weight that we are willing to handle. This will determine the maximum weight of the columns of our LDGM SM code. In this paper, we (somewhat arbitrarily) choose a column weight of 3, corresponding to a maximum stabilizer measurement weight of 12. Because this code is relatively small, there is a significant amount of overlap between randomly chosen stabilizer supports, so only a proportion of the measured stabilizers will have weight 12, and several will have weight 8 or 10. 

    With these choices made, the resulting task is to construct a matrix usable as an SM code which has 24 rows, 60 columns, and a column weight of at most 3. The distance of the overall SM code is no longer tunable: it will be determined by the code construction process, which we now focus on.

\section{Constructing LDGM SM Codes}
\label{sec:ldgm_sm_codes_construction}
The construction of LDGM SM codes is done in a two-step procedure. In the first step, we use the Progressive Edge-Growth (PEG) algorithm \cite{hea05} to design a protograph \cite{tho03}, a relatively small bipartite graph that serves as the base graph of a graph covering. In the second step, the graph covering is constructed using the low-complexity quasi-cyclic PEG algorithm presented in \cite{gfsm23b}. In the following, we explain these two steps in detail.

\subsection{Protograph Design Using PEG}
\label{protograph_design}

The Progressive Edge-Growth (PEG) algorithm \cite{hea05} is a suboptimal tool used to construct the incidence matrix of bipartite graphs with vertices satisfying some prescribed degree distributions. To understand how this algorithm works, consider the following. 

Let $n_c$ and $n_v$ be two positive integers with $n_c<n_v$, and let $B=\left(b_{ij}\right)$ be an $n_c\times n_v$ matrix with nonnegative integer entries. If we take the matrix $B$ as the incidence matrix of a graph, then this graph, denoted by $G_B$, is bipartite. The nodes or vertices of a bipartite graph can be split in two disjoint classes of nodes: check nodes and symbol nodes. Each of the $n_c$ rows of $B$ corresponds to a check node and each of the $n_v$ columns of $B$ corresponds to a symbol node. Let $G_B=(V,E)$ be the bipartite graph obtained from $B$, where $V=V_c\cup V_s$ is the set of nodes, $V_c=\{c_0,c_1,\dots,c_{n_c-1}\}$ is the set of check nodes, and $V_s=\{s_0,s_1\dots,s_{n_v-1}\}$ is the set of symbol nodes. The set $E$, with $E\subseteq V_c\times V_s$, is the set of edges $(c_i,s_j)\in E$ for which $b_{ij}\neq0$, and $0\leq i\leq n_c-1$, $0\leq j\leq n_v-1$. We say that the graph $G_B$ is $(d_s,d_c)$-regular if each symbol node has $d_s$ check nodes as neighbors and each check node has $d_c$ symbol nodes as neighbors; otherwise, we say that the graph is irregular. 

Let $d_{s_j}$ denote the degree of the symbol node $s_j$, $0\leq j\leq n_v-1$, i.e., the number of edges incident to it. After a reordering of the symbol nodes, if necessary, we have the nondecreasing order $d_{s_0}\leq d_{s_1}\leq\cdots\leq d_{s_{n_v-1}}$. Then we denote the symbol degree sequence by $D_s=\{d_{s_0}, d_{s_1},\dots, d_{s_{n_v-1}}\}$. Similarly, let $d_{c_i}$ denote the degree of the check node $c_i$, $0\leq i\leq n_c-1$. After a reordering of the check nodes, if necessary, we have the nondecreasing order $d_{c_0}\leq d_{c_1}\leq\cdots\leq d_{c_{n_c-1}}$. Then we denote the check degree sequence by $D_c=\{d_{c_0}, d_{c_1},\dots, d_{c_{n_c-1}}\}$.

Once a node degree sequence is prescribed, either $D_s$ or $D_c$, the PEG algorithm will start adding ones to an $n_c\times n_v$ all-zero matrix $B$ in such a way that the node degree sequence is satisfied. For the specific values of $n_c$, $n_v$, and either of the node degree sequences, the PEG algorithm will avoid the formation of short cycles in the graph, or delay their formation if unavoidable. For more details about this strategy, we refer the reader to \cite{hea05}.

\begin{remark}
To construct bi-regular graphs using the PEG algorithm, both node degree sequences $D_s$ and $D_c$ should be prescribed. However, the algorithm is not guaranteed to return a bipartite graph with the desired regularity since the values $n_c$ and $n_v$ may not be large enough to allow this. 
\end{remark}

%

The protographs that we construct in this paper not only have a desired symbol node degree sequence $D_s$, but also have desired sizes. To construct the generator matrices of our LDGM SM codes, we first prescribe the number of rows $m$ and columns $n$. Due to the nature of our code design, which involves the construction of graph covers of the protographs, the numbers $m$ and $n$ satisfy the equations $m=n_cN$ and $n=n_vN$ for some positive integers $n_c$, $n_v$, and $N$. For each choice of $N$, we obtain the number of check nodes $n_c$ and symbol nodes $n_v$ that our protograph should have. For example, if $m=24$ and $n=60$, then the possible values for $N$ are the divisors of both 24 and 60, which are 1, 2, 3, 4, 6, and 12. Hence, for these values of $N$, the protographs can be described by matrices with dimensions $24\times60$, $12\times30$, $8\times20$, $6\times15$, $4\times10$, and $2\times5$, respectively. Consider the following example.
\begin{example}
\label{example_protographs}
We used the PEG algorithm to construct a few protographs with uniform column weight 3. In other words, all these protographs have $D_s=\{3,3,\dots,3\}$. Consider the following protographs $B_{2,5}$, $B_{4,10}$, $B_{6,15}$, and $B_{8,20}$:
\begin{align*}
&\left[\begin{matrix}1&2&1&2&2 \\ 2&1&2&1&1\end{matrix}\right] , 
\left[\begin{matrix}1&0&1&1&1&0&1&1&0&1 \\ 0&1&1&1&1&1&1&0&1&1 \\ 1&1&0&0&1&1&1&1&1&0 \\ 1&1&1&1&0&1&0&1&1&1\end{matrix}\right], \\
&\left[\begin{matrix}1&1&0&0&1&0&1&1&0&0&0&1&0&1&1 \\ 1&0&0&1&0&0&0&1&1&0&1&1&1&0&0 \\ 1&1&1&0&0&1&0&0&0&1&1&0&0&1&0 \\ 0&1&1&1&1&0&0&1&1&1&0&0&0&0&0 \\ 0&0&0&0&1&1&1&0&1&0&1&0&1&1&1 \\ 0&0&1&1&0&1&1&0&0&1&0&1&1&0&1\end{matrix}\right], 
\end{align*}
\begin{equation*}
\setlength{\arraycolsep}{3.5pt}
\left[\begin{matrix}1&0&0&1&0&0&1&0&0&0&0&1&0&0&0&0&1&0&1&1 \\ 1&1&0&1&0&1&0&0&1&0&1&0&0&0&1&0&0&1&0&0 \\ 0&0&0&0&0&1&0&0&0&0&0&0&1&1&1&1&1&0&0&1 \\ 0&1&1&0&1&0&0&0&0&1&1&0&0&1&0&0&0&0&1&1 \\ 0&1&0&0&1&0&1&0&1&0&1&1&0&0&0&1&0&0&0&0 \\ 0&0&1&1&0&1&0&1&1&1&0&1&0&1&0&0&0&0&0&0 \\ 1&0&1&0&1&0&0&1&0&1&0&0&1&0&0&0&1&1&0&0 \\ 0&0&0&0&0&0&1&1&0&0&0&0&1&0&1&1&0&1&1&0\end{matrix}\right].
\end{equation*}

\end{example}

\subsection{Lifting the protograph}
Let $B={(b_{ij})}_{n_c\times n_v}$ be the incidence matrix of a protograph designed using the strategy discussed in Section \ref{protograph_design}, where $b_{ij}$ is a nonnegative integer for $i\in[n_c]$ and $j\in[n_v]$, and where $[l]=\{0,1,\dots,l-1\}$. From $B$, we construct a matrix $H={(H_{ij})}_{n_c\times n_v}$,  
where each $H_{ij}$, for $i\in[n_c]$ and $j\in[n_v]$, is a summation of $b_{ij}$ $N\times N$ circulant permutation matrices if $b_{ij}$ is nonzero, and with the $N\times N$ all-zero matrix if $b_{ij}=0$. Graphically, this operation is equivalent to taking an $N$-fold graph cover, or \textit{lifting}, of the protograph. We call $N$ the \textit{lifting factor}.

Let $x^r$ denote the $N\times N$ circulant permutation matrix obtained by circularly shifting to the left, by $r$ positions modulo $N$, the entries of the $N\times N$ identity matrix $I$. For simplicity in the notation, let $p_{ij}(x)$ be the polynomial representation of $H_{ij}$, where $p_{ij}(x)=\sum_{l=0}^{N-1}a_{l}x^{l}$ and $a_{l}\in\{0,1\}$ for all $l\in[N]$. Each polynomial $p_{ij}(x)$ has weight $b_{ij}$. Then we can rewrite the matrix $H$, using the polynomial representation, as $H={(p_{ij})}_{n_c\times n_v}$.

We used the low-complexity quasi-cyclic PEG algorithm presented in \cite{gfsm23b} to construct graph covers of the protographs in Example \ref{example_protographs}.

\begin{example}
\label{example_lifting_protographs}

In this example, we lift the protographs constructed in Example \ref{example_protographs} using circulant permutation matrices to obtain scalar matrices of size $24\times60$. From $B_{2\times5}$, consider the matrices $H_{2\times5}^{(1)}$ and $H_{2\times5}^{(2)}$ given by
\begin{equation*}
H_{2\times5}^{(1)}=\left[\begin{matrix}
x^{6} & x^{7} + x^{6} & 1 & x^{6} + x & x^{9} + x^{6} \\
x^{7} + x^{5} & x^{10} & x^{10} + x^{7} & x^{6} & x^{3}
\end{matrix}\right]
\end{equation*}
and
\begin{equation*}
H_{2\times5}^{(2)}=\left[\begin{matrix}
x^{6} & x^{10} + x^{9} & x^{9} & x^{11} + x^{6} & x^{4} + x \\
x^{9} + x^{7} & x^{8} & x^{6} + x^{3} & x^{6} & x^{6}
\end{matrix}\right].
\end{equation*}
For lifting factor $N_{2\times5}=12$, these matrices give codes with parameters $[60,24,7]$. From $B_{4\times10}$, consider the matrices $H_{4\times10}^{(1)}$ and $H_{4\times10}^{(2)}$ given by
\begin{equation*}
H_{4\times10}^{(1)}=\left[\begin{matrix}
x^{5} & 0 & x^{3} & x^{4} & 1 & 0 & x^{3} & x^{3} & 0 & x^{5} \\
0 & x^{4} & x & x & 1 & x^{2} & x^{4} & 0 & x^{2} & x \\
x & x & 0 & 0 & x & 1 & 1 & x & x & 0 \\
x^{2} & 1 & x^{2} & 1 & 0 & x^{2} & 0 & x & x^{5} & x^{5}
\end{matrix}\right]
\end{equation*}
and
\begin{equation*}
H_{4\times10}^{(2)}=\left[\begin{matrix}
1 & 0 & x^{2} & x^{4} & x^{4} & 0 & x & 1 & 0 & x^{2} \\
0 & x^{4} & 1 & x^{3} & x & x^{4} & x^{3} & 0 & x^{4} & x^{3} \\
x^{2} & 1 & 0 & 0 & x^{5} & x & x^{4} & x^{5} & x^{3} & 0 \\
1 & x & 1 & x^{5} & 0 & x^{3} & 0 & x^{5} & x^{2} & x^{4}
\end{matrix}\right].
\end{equation*}
For lifting factor $N_{4\times10}=6$, these matrices give codes with parameters $[60,24,7]$. From $B_{6\times15}$, consider the matrix $H_{6\times15}$ given by
\begin{equation*}\small\setlength{\arraycolsep}{4pt}
\left[\begin{matrix}
1 & 1 & 0 & 0 & x & 0 & x^{3} & 1 & 0 & 0 & 0 & x & 0 & x^{3} & 1 \\
x^{2} & 0 & 0 & x^{2} & 0 & 0 & 0 & x & x^{3} & 0 & 1 & 1 & x & 0 & 0 \\
x^{2} & x^{3} & x^{3} & 0 & 0 & x & 0 & 0 & 0 & x^{3} & x^{3} & 0 & 0 & x^{3} & 0 \\
0 & x^{3} & 1 & 1 & x^{3} & 0 & 0 & x & x^{2} & x & 0 & 0 & 0 & 0 & 0 \\
0 & 0 & 0 & 0 & 1 & 1 & 1 & 0 & x^{2} & 0 & x & 0 & x & x^{3} & x^{2} \\
0 & 0 & x & x^{2} & 0 & 1 & x^{3} & 0 & 0 & 1 & 0 & x^{3} & x^{3} & 0 & x^{3}
\end{matrix}\right].
\end{equation*}
For lifting factor $N_{6\times15}=4$, this matrix gives a code with parameters $[60,24,7]$. From $B_{8\times20}$, consider the matrix $H_{8\times20}$ given by
\begin{equation*}\tiny \setlength{\arraycolsep}{3pt}
\left[\begin{matrix}
x^{2} & 0 & 0 & x^{2} & 0 & 0 & x & 0 & 0 & 0 & 0 & x & 0 & 0 & 0 & 0 & x^{2} & 0 & x & 1 \\
x & x & 0 & 1 & 0 & x^{2} & 0 & 0 & x & 0 & x^{2} & 0 & 0 & 0 & x^{2} & 0 & 0 & x^{2} & 0 & 0 \\
0 & 0 & 0 & 0 & 0 & x^{2} & 0 & 0 & 0 & 0 & 0 & 0 & 1 & x^{2} & 1 & x^{2} & x & 0 & 0 & 1 \\
0 & 1 & 1 & 0 & 1 & 0 & 0 & 0 & 0 & 1 & x^{2} & 0 & 0 & 1 & 0 & 0 & 0 & 0 & 1 & 1 \\
0 & x^{2} & 0 & 0 & x & 0 & x^{2} & 0 & 1 & 0 & x^{2} & x & 0 & 0 & 0 & x^{2} & 0 & 0 & 0 & 0 \\
0 & 0 & x & x^{2} & 0 & 1 & 0 & x & x & 1 & 0 & 1 & 0 & x^{2} & 0 & 0 & 0 & 0 & 0 & 0 \\
1 & 0 & 1 & 0 & x^{2} & 0 & 0 & x & 0 & x & 0 & 0 & x & 0 & 0 & 0 & x & x^{2} & 0 & 0 \\
0 & 0 & 0 & 0 & 0 & 0 & 1 & 1 & 0 & 0 & 0 & 0 & x & 0 & x^{2} & x^{2} & 0 & 1 & x & 0
\end{matrix}\right].
\end{equation*}
For lifting factor $N_{8\times20}=3$, this matrix gives a code with parameters $[60,24,7]$.
\end{example}

\subsection{Resulting SM Codes}

\begin{figure}[htbp]
\label{fig:}
    \centering
    \begin{subfigure}[b]{0.44\textwidth}
        \centering
        \includegraphics[width=\textwidth]{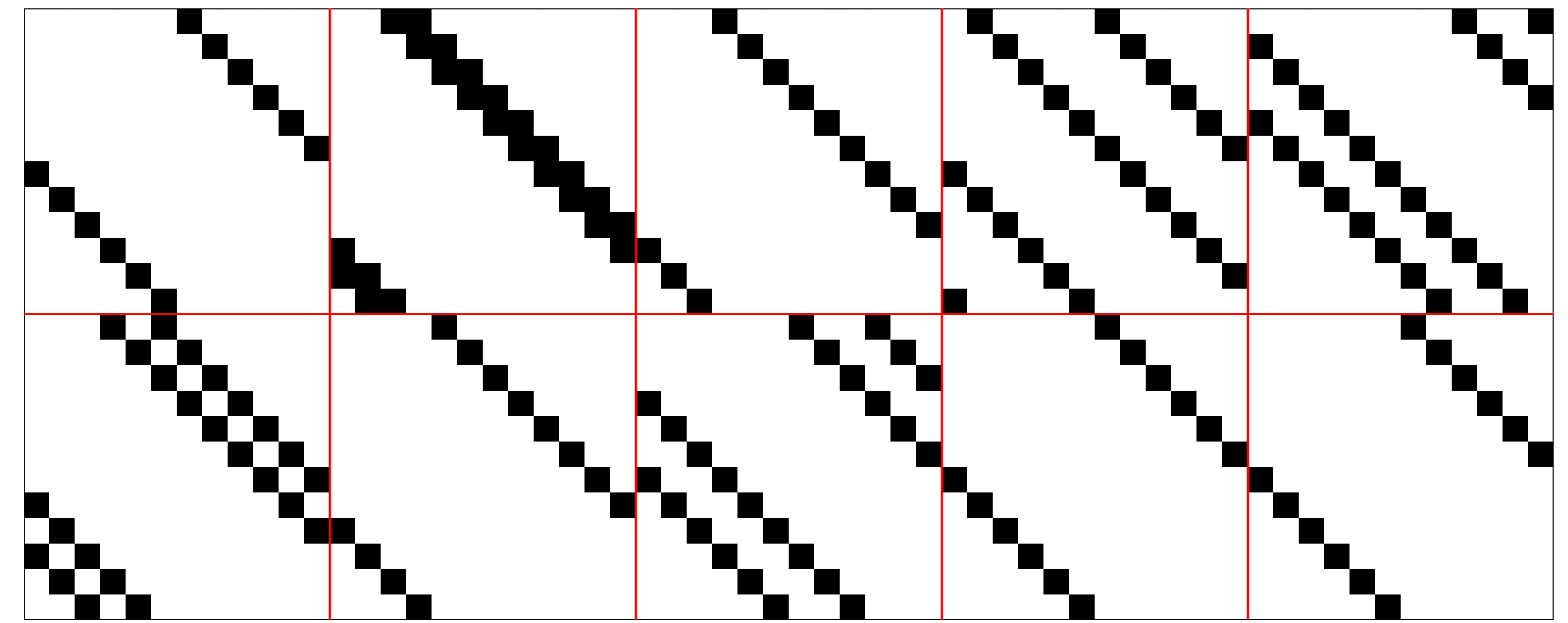}
        \caption{The code generated from $H_{2\times 5}^{(2)}$}
        \label{fig:subfig_a}
    \end{subfigure}%
    \hfill%
    \begin{subfigure}[b]{0.44\textwidth}
        \centering
        \includegraphics[width=\textwidth]{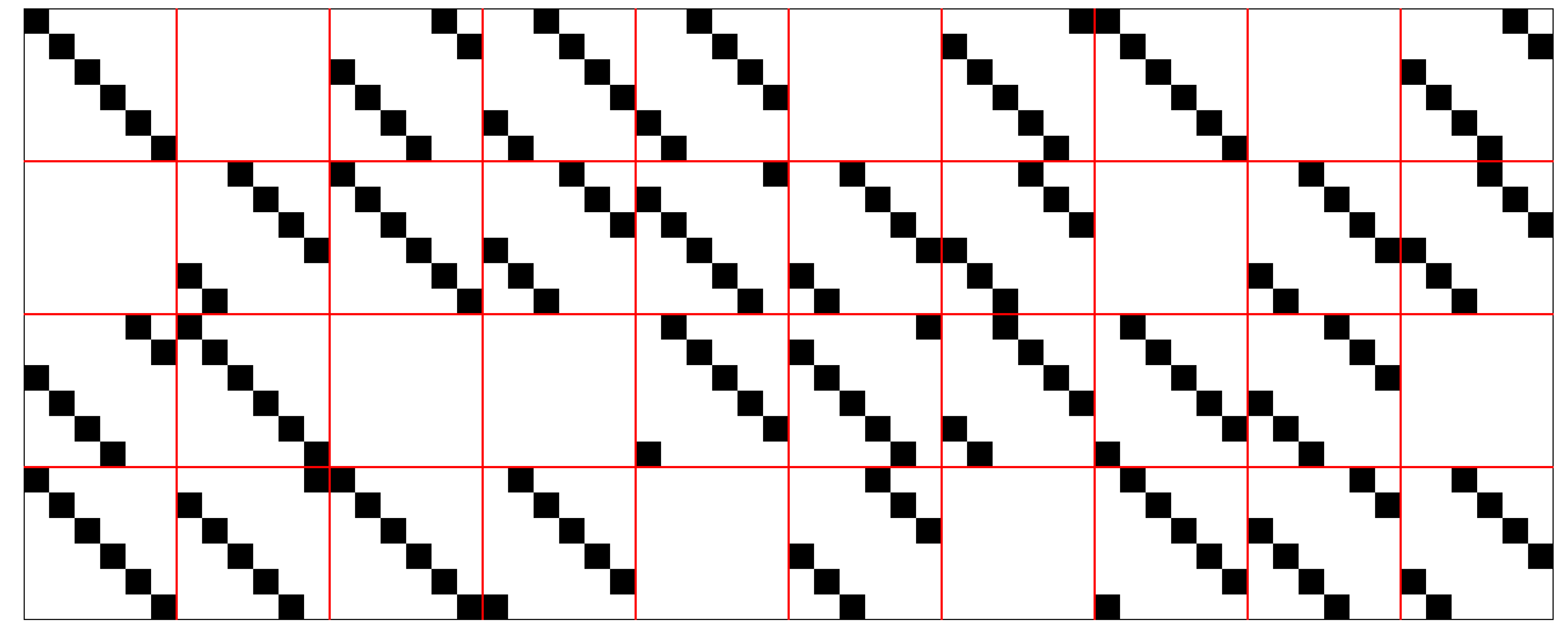}
        \caption{The code generated from $H_{4\times 10}^{(2)}$}
        \label{fig:subfig_b}
    \end{subfigure}
    
    \medskip
    
    \begin{subfigure}[b]{0.44\textwidth}
        \centering
        \includegraphics[width=\textwidth]{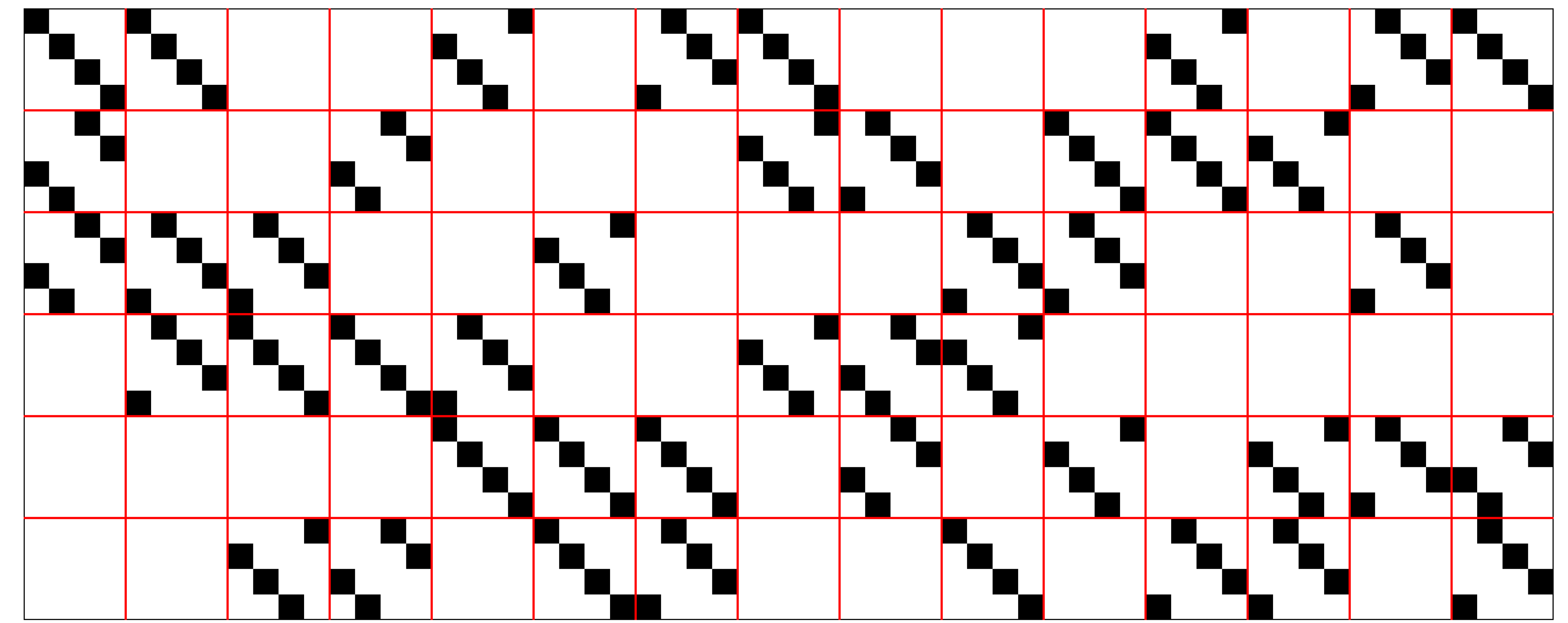}
        \caption{The code generated from $H_{6\times 15}$}
        \label{fig:subfig_c}
    \end{subfigure}%
    \hfill%
    \begin{subfigure}[b]{0.44\textwidth}
        \centering
        \includegraphics[width=\textwidth]{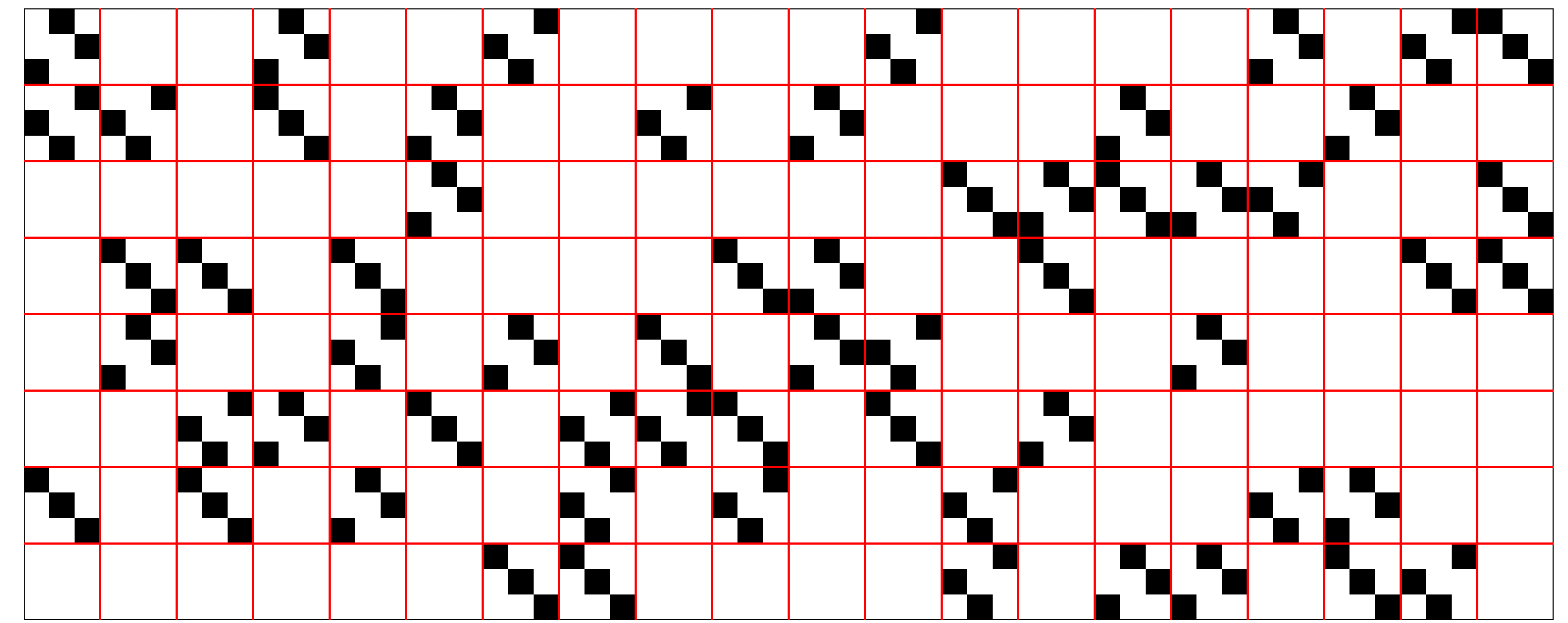}
        \caption{The code generated from $H_{8\times 20}$}
        \label{fig:subfig_d}
    \end{subfigure}
    \caption{Visualizations of the codes generated from the protograph matrices. White squares represent a 0 in the corresponding location, while black squares represent a 1. The red lines serve to divide the matrix into squares, where each square corresponds to an entry in the corresponding protograph. For example, the top left square in graph (a) corresponds to the entry $x^6$ in $H_{2\times 5}^{(2)}$, and shows the corresponding cyclic permutation. Similarly, in (c), the top-left square is the $4\times4$ identity matrix, which corresponds to the top-left entry $1$ in the protograph.}
    \label{fig:code_visualizations}
\end{figure}

The LDGM SM codes resulting from our constructions are all distance-7 codes, and therefore able to correct against up to 3 errors on the measurements. This is an improvement over the 5-fold repeated measurements that would be standard, and getting similar performance out of repeated syndrome measurement would require 7-fold repetition, involving a total of 168 stabilizer measurements. Compared to this, the LDGM encoding represents an improvement by a factor of 2.8. In Fig. \ref{fig:code_visualizations}, we present the visualization of some of the generator matrices constructed in Example \ref{example_lifting_protographs}.
\section{Constraints on generated codes}

There is an important constraint on the amount of improvement that we can see from such an LDGM SM code compared to the repetition code. 

If a binary encoding matrix has $R$ rows and $C$ columns, with each column having constant weight $w_C$, then the total number of nonzero entries is given by $Cw_C$. Then these $Cw_C$ nonzero entries are split up among the $R$ rows, giving an average row weight of $\frac{Cw_C}{R}$ and a maximum Hamming distance of \begin{equation}
\label{eq:d_max}
d_{max}=\left\lfloor\frac{Cw_C}{R}\right\rfloor.
\end{equation} 

In the case of a distance-$d$ QLDPC code with $\ell$ stabilizer generators, the encoding matrix corresponding to repeated syndrome extraction has $\ell$ rows and $d\ell$ columns. If we want to perform syndrome extraction using an LDGM code with $\ell$ rows and $C$ columns, we can characterize the improvement factor between these two matrices as the ratio of their number of columns, which we will denote $\alpha=\frac{C}{d\ell}$. Note that $\alpha\in(0,1]$. Then we can denote the number of columns $C$ as $d\ell \alpha$, and the total number of nonzero entries in the LDGM SM encoding matrix as the product of this number and the column weight: $w_Cd\ell\alpha$. Finding the maximum Hamming distance of this LDGM code using \eqref{eq:d_max}, and recalling that the number of rows in this matrix is $\ell$, we get
\begin{equation*}
d_{max} \leq  \frac{w_Cd\ell\alpha}{\ell}=w_Cd\alpha.
\end{equation*}

Since we want to maintain global fault tolerance, we want $d_{max}$ to be at least equal to the distance $d$ of the code. This gives us the requirement that $d_{max}\geq d$. Combined with the previous equation, this gives us the constraint that
\begin{equation*}
d\leq w_Cd\alpha\implies 1\leq w_C\alpha\implies\alpha\geq \frac{1}{w_C}.
\end{equation*}
This then means that the improvement factor of these codes is limited by the acceptable weight of the columns; in a scheme where we can handle the product of at most 5 stabilizer generators, we cannot perform fault-tolerant syndrome measurements in fewer than $\frac15$ as many measurements as repeated syndrome extraction for the same distance.

From this analysis, it follows that syndrome measurement schemes that preserve the QLDPC property cannot be performed in fewer than $O(d\ell)$ total measurements, using $\frac{d\ell}{w_C} -\ell = \ell \left(\frac{d}{w_C}-1\right)\in O(d\ell)$ additional measurements. We often use the number of additional required measurements as a proxy for the improvement of syndrome measurement code schemes over repeated syndrome extraction as in \cite{Guttentag23}. This implies that this style of syndrome measurement code cannot asymptotically improve over repeated syndrome extraction.

\section{Modeling and Results}

To analyze the behavior of these LDGM codes as SM codes, we compare their behavior under phenomenological noise models. In particular, we focus on a model where noise occurs on the syndrome qubits with a significantly higher probability than on the data qubits. We assume that all syndrome bits have an equal probability of experiencing a measurement error; the fact that most measured stabilizers have a similar weight and therefore involve a similar number of gates allows this assumption to be largely correct.

\subsection{Importance sampling} 

A common obstacle in the Monte Carlo simulation of error rates is that as the probability of a particular error $p$ decreases, the vast majority of traditionally generated errors will be zero-weight (trivial) errors. In fact, for a code with distance $d=2t+1$, the lowest-weight error that can cause a logical error will be one of weight $t+1$, which would have a probability $p^{t+1}$ of occurring \cite{Gutierrez2019}. We would expect to need to perform $\frac{1}{p^{t+1}}$ trials to see even one such error, and would spend almost all of our computational effort on situations with no resulting errors. 

To deal with this, we use a technique known as \textit{importance sampling} \cite{Gutierrez2019,Guttentag23}. In this technique, we intentionally generate errors of a certain weight $w$ (say, weight 4); the random step would then involve randomly sampling across the space of weight-4 errors, under the assumption that such errors are uniformly distributed. 

We perform a Monte Carlo estimate via repeated sampling over this space to determine the likelihood of such an error resulting in a logical error; we denote this probability to be $p_L(w)$ for a weight-$w$ error. We then estimate the probability of a weight-$w$ error ocurring by the binomial theorem; for a probability of errors on an individual location given as $p$, and $n$ total locations, the probability of exactly $w$ errors occurring, $A_w(p)$, is given by
\begin{equation*}
A_w(p)=\binom{n}{w}p^w(1-p)^{n-w}.
\end{equation*}
We can therefore estimate the probability of a logical error occurring (given the probability $p$ of an error on each individual site), $Pr(p)$. We find this probability by taking a weighted sum over error weights $w$, where the probability $p_L(w)$ of a weight-$w$ error causing a logical error is weighted by the probability of that error occurring, $A_w(p)$: that is,
\begin{equation*}
Pr(p) = \sum_{w}p_L(w)A_w(p).
\end{equation*}
For a more complete treatment of the importance sampling methodology, as well as a fuller discussion of how we use this method to handle multiple error types (namely, both data qubit and ancilla qubit errors), see \cite{Guttentag23,Gutierrez2019}.

The process of simulation involves the following steps. First, we generate a lookup decoder for the LDGM SM code being considered. This is done as a preprocessing step, as it is very computationally intensive. Second, we encode the parity-check matrix of the underlying quantum code using the LDGM code, following the method outlined in Section \ref{subsec:protecting_syndrome_information}. This results in a set of 60 stabilizer group elements to be measured. Third, we measure these 60 stabilizers, obtaining a length-60 binary syndrome. Fourth, we decode this syndrome using the lookup decoder we developed for the LDGM SM code; the output of this is a length-24 syndrome. Fifth, we use this length-24 syndrome to determine the most likely qubit error to occur and perform a correction.

 \begin{figure}[th]
    \centering
    \includegraphics[width=\linewidth]{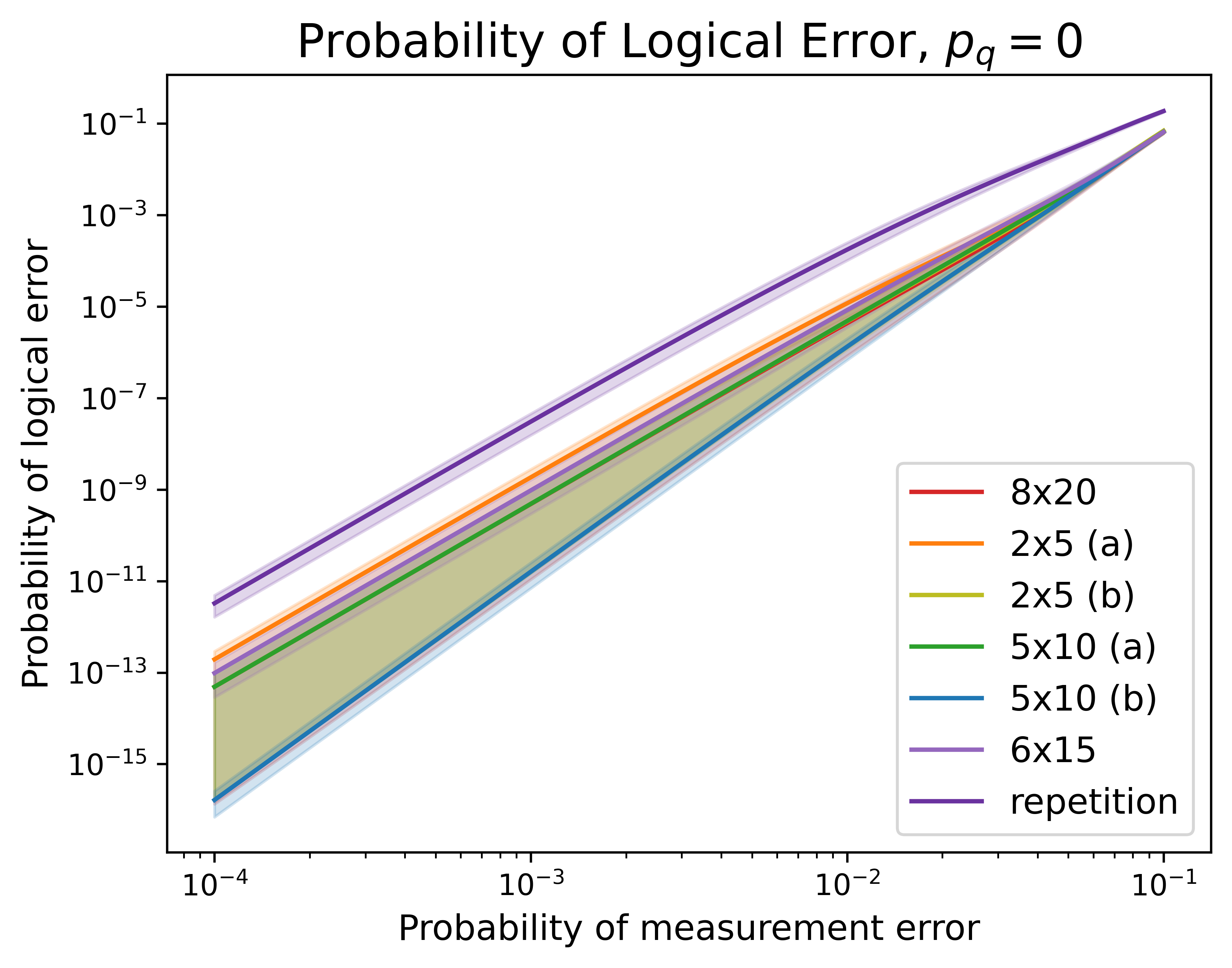}
    \caption{A comparison between the probability of logical error resulting from measurement error for the codes generated by the six different protographs, as well as a comparison to 5-fold repetition. This only considers errors manifesting as bit-flips on the measurements, and not qubit errors.}
    \label{fig:results1}
\end{figure}

 \begin{figure}[th]
    \centering
    \includegraphics[width=\linewidth]{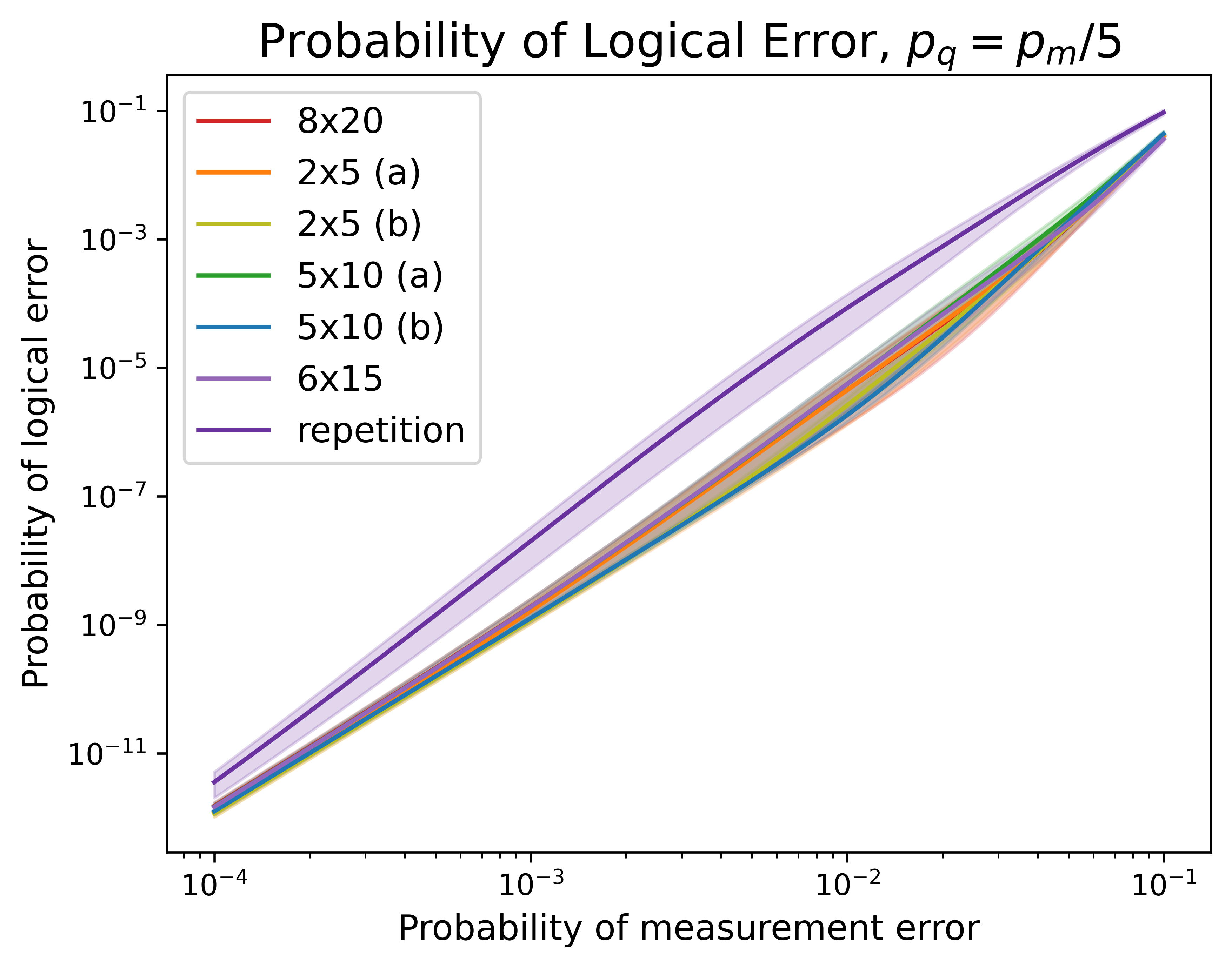}
    \caption{A comparison between the probability of logical error resulting from measurement and qubit error for the codes generated by the six different protographs, as well as a comparison to 5-fold repetition. This assumes errors occur with probability $p_m$ on each stabilizer measurement, and qubit errors occur with probability $p_q=p_m/5$ on all qubits.}
    \label{fig:results2}
\end{figure}

\subsection{Results}
To ascertain the performance of these LDGM SM codes, we compare them both together with the $[120,24,5]$ 5-fold repetition code. We will refer to Fig. \ref{fig:results1} and Fig. \ref{fig:results2} when discussing these results. 

It is notable that the margins of error on these results are such that all of the LDGM codes act indistinguishably, while all being significantly more effective at performing fault-tolerant syndrome extraction than the repetition code. The reasons for this are twofold. First, all the LDGM codes have distance 7, which means that they can correct a larger number of syndrome measurement errors than the repetition code, which has distance 5. The constructed LDGM SM codes only involving 60 total measurements confers another advantage, as the probability of a high-weight error occurring is lower on fewer sites. 

There are two sets of results. Those shown in Fig. \ref{fig:results1} are under a phenomenological model in which we consider only measurement errors; this model allows for the distance effects to dominate. The results in Fig. \ref{fig:results2} are under a phenomnoloical model in which we consider both syndrome and qubit errors. As the probability of measurement error decreases, the qubit errors begin to dominate, which results in the behavior of the 5-fold repetition code and the LDGM codes being similar at lower error probabilities. Nonetheless, these results suggest that in a regime where measurement errors are more likely than qubit errors, there are measurement error probabilities where a distinct advantage is conferred from the use of such LDGM SM codes instead of simple repetition codes.

It is worth noting that the behavior of this method under circuit-level noise may be different, and it is necessary to perform further simulations under a more detailed noise model in order to successfully determine the ways in which syndrome measurement errors and data-qubit errors combine to affect the probability of logical error. It may be possible to implement the techniques described by Lin et. al. towards this end \cite{lin2025dynamiclocalsingleshotchecks}.

\subsection{Conclusions}
From the results, we conclude that constructing LDGM SM codes with low-weight columns allows us to perform syndrome extraction fault-tolerantly while performing fewer total measurements than are required to maintiain the same global fault-tolerant behavior. This also allows us to guarantee that no high-weight stabilizer operators are measured, which allows us to be confident that the measurements performed do not spread errors to arbitrarily many locations on the qubits. In fact, this methodology allows us to have complete control over the maximum weight of performed measurements by controlling the column weights. 

We also conjecture that this allows us to maintain the single-shot property discussed in \cite{Bombin2015} for certain codes. The core concept that allows for the single-shot property is \textit{confinement}, which numerically quantifies the amount by which errors can spread during the error correction process. While this does not exhibit \textit{geometric} confinement, as qubits from geometrically distant locations on the surface code grid are involved in the same qubit measurement, it does exhibit weight confinement.

\subsection{Further work}

This project is primarily intended to show that it is indeed possible to preserve the QLDPC property of a quantum code while performing fault-tolerant syndrome measurements in a significantly shorter amount of time. While we have shown that this process does theoretically preserve the QLDPC property, it remains to be shown that this encoding does perform as expected when simulated at the circuit level. Additionally, significant work remains to be done to marry the frameworks of single-shot \cite{Bombin2015} and data-syndrome \cite{Ashikhmin2020,Guttentag23,Fujiwara2014} codes. Data-syndrome code schemes that may preserve the single-shot property, such as those discussed in this work, provide a jumping-off point from which the connections between these two formalisms can be explored.

Additional areas of improvement with this method are also possible, by attempting to optimize or preserve other properties. While this methodology requires fewer overall syndrome measurements, it does potentially involve a larger number of two-qubit gates to be performed, so it may be ideal to try to shorten the number of measurements further. Alternatively, it may be desirable to guarantee that the stabilizer group elements being measured have supports that are geometrically local; currently this is not guaranteed by the construction of these LDGM SM codes. The study of data-syndrome and syndrome measurement codes is ongoing and exciting, and these questions will be explored in the future.

\section{Acknowledgments}

This work was supported by the National Science Foundation (NSF) award PHY-2514847 and by the Office of the Director of National Intelligence (ODNI), Intelligence Advanced Research Projects Activity (IARPA), under the Entangled Logical Qubits program through Cooperative Agreement Number W911NF-23-2-0216.

\nopagebreak	
\thispagestyle{empty}
\printbibliography
\balance

\end{document}